\documentclass[3p,final,times]{elsarticle}

\usepackage{moresize}
\usepackage{amsmath}
\usepackage{amsfonts}
\usepackage{bm}
\usepackage{lipsum}
\usepackage{enumerate}
\usepackage{graphicx}
\usepackage{multirow}
\usepackage{amsfonts}
\usepackage{subcaption}

\usepackage{mathrsfs}
\usepackage{epstopdf}
\usepackage{url}
\usepackage{verbatim}
\usepackage{amssymb}
\usepackage{hyperref}
\usepackage{color}
\usepackage{listings}
\usepackage{caption}
\usepackage{subcaption}
\usepackage{diagbox}

\def\@begintheorem#1#2{\par\bgroup{\scshape #1\ #2. }\it\ignorespaces}
\def\@opargbegintheorem#1#2#3{\par\bgroup%
   {\scshape #1\ #2\ ({\upshape #3}). }\it\ignorespaces}
\def\@endtheorem{\egroup}

\if@onethmnum
  \newtheorem{theorem}{Theorem}
  \newtheorem{lemma}[theorem]{Lemma}
  \newtheorem{corollary}[theorem]{Corollary}
  \newtheorem{proposition}[theorem]{Proposition}
  \newtheorem{definition}[theorem]{Definition}
\newtheorem{example}[theorem]{Example}
\newtheorem{remark}[theorem]{Remark}
\newtheorem{homework}[theorem]{Homework}
\newtheorem{case}[theorem]{}

\else

\fi

\usepackage[T1]{fontenc}

\journal{Applied Mathematics Letters}

\begin{document}

\begin{frontmatter}



\title{A comparative study of dynamic models for gravity-driven particle-laden flows}

\author[1]{\small Wing Pok Lee\fnref{fn1} }
\author[1]{\small Jonathan D. Woo\fnref{fn1}}
\author[2]{\small Luke F. Triplett\fnref{fn1}} 
\author[1]{\small Yifan Gu\fnref{fn1}}
\author[1]{\small Sarah C. Burnett}
\author[1]{\small Lingyun Ding\corref{cor1}}
 \ead{dingly@g.ucla.edu}
\cortext[cor1]{Corresponding author}

\author[1,3]{\small Andrea L. Bertozzi}
\address[1]{\small Department of Mathematics, University of California Los Angeles, Los Angeles, CA, 90095, United States}
\address[2]{\small Department of Mathematics, Duke University, Durham, NC, 27708, United States}
\address[3]{\small Department of Mechanical and Aerospace Engineering, University of California Los Angeles, Los Angeles, CA, 90095, United States}

\fntext[fn1]{Equal contribution}

\begin{abstract}
\small The dynamics of viscous thin-film particle-laden flows down inclined surfaces are commonly modeled with one of two approaches: a diffusive flux model or a suspension balance model. The diffusive flux model assumes that the particles migrate via a diffusive flux induced by gradients in both the particle concentration and the effective suspension viscosity. The suspension balance model introduces non-Newtonian bulk stress with shear-induced normal stresses, the gradients of which cause particle migration. Both models have appeared in the literature of particle-laden flow with virtually no comparison between the two models. For particle-laden viscous flow on an incline, in a thin-film geometry, one can use lubrication theory to derive a compact dynamic model in the form of a $2\times 2$ system of conservation laws. We can then directly compare the two theories side by side by looking at similarities and differences in the flux functions for the conservation laws, and in exact and numerical simulations of the equations. We compare the flux profiles over a range of parameters, showing fairly good agreement between the models, with the biggest difference involving the behavior at the free surface. We also consider less dense suspensions at lower inclination angles where the dynamics involve two shock waves that can be clearly measured in experiments. In this context the solutions differ by no more than about 10\%, suggesting that either model could be used for this configuration.
\end{abstract}


\end{frontmatter}

\section{Introduction}\label{intro}
\small The study of the dynamics of particles in viscous liquids is important due to their wide-ranging applications in fields involving the transport of suspensions, such as in the food industry \cite{chocolate} and in geology \cite{landslides1}. 
However, there is still a need for a fundamental understanding of the physical phenomena occurring in creeping suspensions, including gravitational settling and shear-induced resuspension \cite{guazzelli2018rheology}. 

The present work focuses on the dynamics of particle-laden gravity-driven thin-film flows down an incline. These flows can be captured through simple experiments \cite{murisic2011particle,murisic2013dynamics} by mixing negatively buoyant particles (e.g., glass or ceramic beads) in a viscous liquid (e.g., silicone oil) to form a uniformly mixed suspension, before pouring the mixture into a wide, inclined channel and letting it flow under the influence of gravity. There are two distinct classes of continuum models used to describe particle-laden flows: diffusive flux models (DFMs) and suspension balance models (SBMs). While diffusive flux models describe the flux of the particles in the suspension in terms of the gradients of the particle concentration and the shear rate \cite{murisic2013dynamics,leighton1987shear,Phillips}, suspension balance models relate the migration flux of the particles within the suspension rheology based on volume-averaged balances of mass and momentum for the liquid and particle phases \cite{wong2019fast,morris_boulay,Nott_Brady_1994,nott2011suspension,timberlake2005particle}. 

Both models have been successful in capturing experimental observations of the viscous thin-film flow down an incline \cite{murisic2013dynamics, wong2019fast, murisic2011particle}. More broadly, these models apply to various particle-laden flows, including Couette flows \cite{Phillips, morris_boulay, boyer2011dense}, laminar pipe flow \cite{zhang1994viscous, miller2006normal}, and parallel plate flows \cite{merhi2005particle, xu2016particle}, among others.
The SBM features a non-Newtonian suspension stress with shear-induced normal stresses, which drive particle migration through their gradients. This model includes physically-derived formulas for normal and suspension viscosities and stresses, making it more suitable for solving stress-related problems, particularly in studying the viscoelastic properties of highly-concentrated suspensions \cite{boyer2011dense, xu2016particle}.

We compare the two models directly by examining numerical simulations of vertical equilibrium profiles where suspensions with local concentrations above and below a critical concentration will have different behaviors. We validate numerical simulations with physical experimental results of the thin-film flow in a settled regime. That is, at lower particle concentrations and small inclination angles, when the particles settle to the bottom and the clear liquid advances past the particles at the front of the mixture (e.g. Fig.~\ref{fig:diagram}). We find the suspension balance model predicts a larger fluid velocity than the diffusive flux model, but overall they produce similar predictions for the fluid front and height profile for the suspension. 

This paper is organized as follows. Sections \ref{diffusive flux gov} and \ref{suspension_balance_gov} describe the governing equations of the two models. Section \ref{theoretical_sol} discusses a lubrication approximation to derive models for particle and liquid transport, with a correction to the non-local shear rate. Sections \ref{equilibrium_solutions} and \ref{numerical_sol} include numerical simulations to examine the differences between the transport models, including the equilibrium profiles and the solutions to conservation law equations, respectively. In Sec.~\ref{numerical_sol}, we examine the validity of the models by comparing simulations and experiment. Section \ref{conclusion} documents the conclusions of this paper and future avenues of investigation. 

\section{Governing Equations}\label{gov_eqns}
\begin{figure} 
 \centering
 \includegraphics[width=0.4\linewidth]{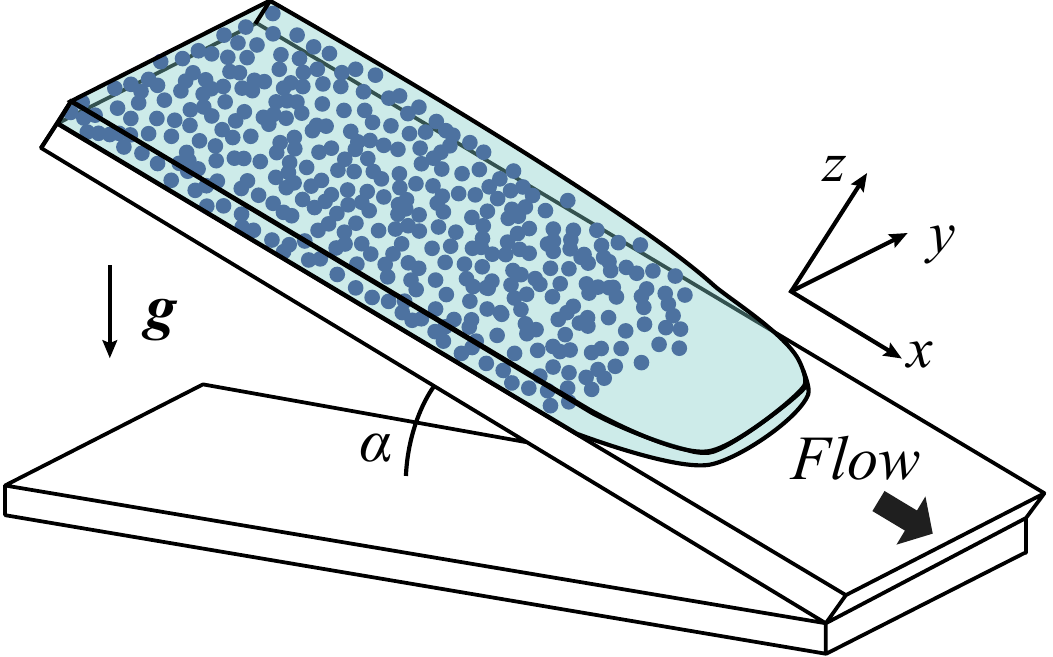}
 \hfill
 \caption[]
 { A schematic of a settled particle-laden suspension on an inclined flat plane. The channel is inclined at an angle $\alpha$ and the flow is driven by gravity indicated by $\mathbf{g}$.}
\label{fig:diagram}
\end{figure}
A schematic of the problem setup is provided in Fig.~\ref{fig:diagram}. We consider two spatial dimensions in the $x$-direction (parallel to the incline) and the $z$-direction (perpendicular to the incline). We assume no variation in the $y$-direction, reducing the problem to a two-dimensional system. In Sections \ref{diffusive flux gov} and \ref{suspension_balance_gov}, we will use this coordinate system to establish the governing equations of the diffusive flux and suspension balance models. We denote $\phi=\phi(t,x,z)$ to be the particle volume fraction and $\mathbf{u}(t,x,z)=(u(t,x,z),w(t,x,z))^T$ to be the particle-liquid mixture velocity.

\subsection{Diffusive Flux Model}\label{diffusive flux gov}
Based on the diffusive flux phenomenology \cite{murisic2013dynamics,leighton1987shear,Phillips}, the particle-fluid mixture is modeled as an incompressible Newtonian fluid, with the density and viscosity modified by the presence of particles. The density of the suspension is $\rho(\phi)=\rho_{\ell}+(\rho_p-\rho_{\ell})\phi$, where $\rho_p$ and $\rho_{\ell}$ are the particle and liquid densities. The viscosity of the suspension is given by Krieger-Dougherty relation \cite{krieger1959mechanism}, $\mu(\phi)=\mu_{\ell}\left(1-\frac{\phi}{\phi_m}\right)^{-2}$, where $\mu_{\ell}$ is the fluid dynamic viscosity and $\phi_m$ is the maximum packing fraction of the particles. Following the literature \cite{murisic2013dynamics}, we choose $\phi_m=0.61$ in this work. Equations for the momentum balance of the suspension (Stokes), the incompressibility, and the particle transport are listed as follows:
\begin{equation} \label{eqn:gov1}
 0=-\nabla p+\nabla\cdot \mathbf{\Sigma}_d+\rho(\phi)\mathbf{g},\quad \nabla \cdot \mathbf{u} = 0,\quad\frac{\partial\phi}{\partial t}+\nabla\cdot (\mathbf{u}\phi+\mathbf{J}_d)=0. 
\end{equation}
Pressure is denoted as $p$ and the gravitational vector is $\mathbf{g}=(g\sin\alpha, -g\cos\alpha)^T$, where $g$ is the gravity acceleration constant and $\alpha$ is the inclination angle relative to the horizontal plane. The suspension stress for the diffusive flux model $\Sigma_d$ is given by $\Sigma_d=2\mu(\phi)\textbf{E}$, where $\textbf{E}=\frac{1}{2}(\nabla \textbf{u}+\nabla \textbf{u}^T)$ is the strain rate tensor. The flux term $\mathbf{J}_d$ in equation~\eqref{eqn:gov1} describes the particle gravitational settling and shear induced migration, which is given by
\begin{equation}
\begin{split}
\label{flux}
\mathbf{J}_d=\frac{d^2\phi(\rho_p-\rho_\ell)\Phi(\phi)}{18\mu(\phi)}\mathbf{g}-\frac{K_cd^2}{4}(\phi^2\nabla\dot{\gamma}+\phi\dot{\gamma}\nabla\phi) - \frac{K_v d^2}{4 \mu(\phi)}\phi^2 \dot{\gamma} \frac{d\mu(\phi)}{d\phi} \nabla \phi,
\end{split}
\end{equation}
where $d$ is the particle diameter, $\Phi(\phi)=1-\phi$ is the hindrance function, and $\dot{\gamma}=\|\frac{\nabla \textbf{u}+\nabla \textbf{u}^T}{2}\|_{F}$ is the shear rate ($F$ denotes the Frobenius norm). $K_c=0.41$ and $K_v=0.62$ are empirical constants. 

\subsection{Suspension Balance Model}\label{suspension_balance_gov}
We adopt the suspension balance model described in \cite{wong2019fast,timberlake2005particle,miller2006normal}. The formation the suspension balance model is similar to the system (\ref{eqn:gov1}), differing only in two ways. First, the suspension stress term for the suspension balance model $\mathbf{\Sigma}_s$ is modified to account for normal stress $\mathbf{\Sigma}_N$ in the suspension. The expression for $\mathbf{\Sigma}_s$ is given by $\mathbf{\Sigma}_s=2\mu^{(s)}(\phi)\textbf{E}+\mathbf{\Sigma}_N$, where the suspension viscosity $\mu^{(s)}(\phi)$ is given by 
\begin{equation}\label{eqn:sbm_visc}
\mu^{(s)}(\phi)=\mu_{\ell}\left(1+\frac{5}{2}\frac{\phi}{(1-\phi/\phi_m)}+I(\phi)\frac{(\phi/\phi_m)^2}{(1-\phi/\phi_m)^2}\right), \quad I(\phi)=m_1+\frac{m_2-m_1}{1+I_0\phi^2/(\phi_m-\phi)^2},
\end{equation}
where the constants used here are $I_0=0.005, m_1=0.32, m_2=0.7$. The normal stress $\Sigma_N$ is given by 
\begin{equation} \label{eqn:shear_rate}
\mathbf{\Sigma}_N=-\mu^{(n)}(\phi)\dot{\gamma}(\Lambda_1\mathbf{e}_x\mathbf{e}_x^T+\Lambda_2\mathbf{e}_z\mathbf{e}_z^T), \quad \mu^{(n)}(\phi)=K_n\mu_{\ell}(\phi/\phi_m)^2/(1-\phi/\phi_m)^2
\end{equation}
where $\mu^{(n)}(\phi)$ is the normal viscosity, and
the constants used are $\Lambda_1=1,\Lambda_2=0.8,K_n=1$, and $\mathbf{e}_x$, $\mathbf{e}_z$ are unit column vectors in the $x$ and $z$-directions, respectively. The second difference is that the particle transport equation consists of a flux $\textbf{J}_s$ which includes the contribution due to particle phase stress $\mathbf{\Sigma}_p$:
\begin{equation}\label{eqn:particle_flux_sbm}
\mathbf{J}_s=\frac{d^2}{18\mu_{\ell}}\Phi(\phi)(\nabla \cdot \mathbf{\Sigma}_p+(\rho_p-\rho_{\ell})\phi \mathbf{g}), \quad \mathbf{\Sigma}_p=(\mu^{(s)}(\phi)-\mu_{\ell})\textbf{E}+\mathbf{\Sigma}_N.
\end{equation}

\subsection{Lubrication Approximation} \label{theoretical_sol}
In most experiments, the characteristic fluid layer thickness $H$ is small compared to the characteristic length $L$ in the direction along the incline, namely $\varepsilon=H/L\ll 1$. 
We use the lubrication approximation technique \cite{oron1997long,ding2023diffusion} to reduce the governing equations of both models described in Sec.~\ref{gov_eqns} to simplified conservation laws for the particle concentration and the film height. We use the following non-dimensional quantities (denoted with a hat) and the corresponding characteristic values for both the diffusive flux and suspension balance models:
\begin{align}
&(x,z)=\left(L\Hat{x},\Hat{z}\right), \quad (u,w)=U(\Hat{u},\varepsilon \Hat{w}), \quad \varepsilon=\frac{H}{L}, \quad U=\frac{H^2\rho_{\ell}g \sin \alpha}{\mu_{\ell}},\quad t=\frac{L}{U}\Hat{t}, \label{eqn:scales_1} \\
&(J_x,J_z)=\left(\varepsilon \frac{d^2U}{H^2}\Hat{J}_x, \frac{d^2U}{H^2}\Hat{J}_z\right), \quad \mu=\mu_\ell\Hat{\mu}, \quad  \mu^{(s)}=\mu_{\ell}\Hat{\mu}^{(s)}, \quad \mu^{(n)}=\mu_{\ell}\Hat{\mu}^{(n)}, \quad p=\frac{U\mu_{\ell}}{H}\Hat{p}.\label{eqn:scales_2}
\end{align}
To ensure a continuum model, we assume the particle diameter is much smaller than the fluid depth, $ (d/H)^2=\varepsilon^{\beta}\ll 1$, where $0<\beta<1$, and otherwise assume $d$ large enough so that the particles are non-colloidal.
In the limit $\varepsilon\rightarrow \infty$, $\dot{\gamma} \approx \left|\frac{\partial u}{\partial z} \right|+ \mathcal{O} (\varepsilon)$ and the standard asymptotic calculation gives the following equation that governs the leading order approximation of the shear stress $\sigma=\mu(\phi)\frac{\partial u}{\partial z}$ and particle volume fraction $\phi$ for the diffusive flux model
\begin{align}\label{eqn:dfm_ode}
  \partial_{s}\Tilde{\sigma}+\left(1+\frac{\rho_p-\rho_{\ell}}{\rho_{\ell}}\Tilde{\phi}\right)=0, \quad 
\left(1+\frac{2(K_v-K_c)}{K_c}\frac{\tilde{\phi}}{\phi_m-\tilde{\phi}}\right)\tilde{\sigma}\partial_{s}\tilde{\phi}+B_d-(B_d+1)\tilde{\phi}-\frac{\rho_p-\rho_\ell}{\rho_\ell}\tilde{\phi}^2=0,\quad \tilde{\sigma}(1)=0,
\end{align}
where $B_d=2(\rho_p-\rho_\ell)\cot\alpha/9\rho_\ell K_c$ is a non-dimensional buoyancy parameter measuring the strength of settling due to gravity in the $z$-direction relative to the strength of shear-induced migration.
The first equation in \eqref{eqn:dfm_ode} comes from the approximation of Stokes equation, and the second equation comes from the particle transport equation. In \eqref{eqn:dfm_ode}, we have defined $\tilde{\phi}$ and $\tilde{\sigma}$ with normalized height $s=z/h$ ($h$ is the height of the suspension) by
\begin{equation}
\Tilde{\phi}(t, x;s):=\phi(t, x; h(t, x)s)=\phi(t, x;z), \quad 
\Tilde{u}(t, x;s):=\frac{u(t, x; h(t, x)s)}{h(t,x)^2}, \quad \Tilde{\sigma}(t, x;s) :=\frac{ \sigma (t, x; h(t, x)s)}{h(t, x)}.
\end{equation}
The solution of \eqref{eqn:dfm_ode} is uniquely determined given  $\phi_0 (x,t) =\int_0^1\tilde{\phi} (t,x;s)\;ds $. Therefore, the dependence of $\phi$ on $x$ and $t$ is captured though $\phi_{0}$. Using a similar procedure, to leading order in $\varepsilon$, the equilibrium model for the suspension balance model is given by 
\begin{equation} \label{eqn:SBM_eq}
\partial_{s}\Tilde{\sigma}+\left(1+\frac{\rho_p-\rho_{\ell}}{\rho_{\ell}}\Tilde{\phi}\right)=0, \quad \partial_{s} (R(\Tilde{\phi})\Tilde{\sigma})+B_s \Tilde{\phi} = 0, \quad \tilde{\sigma}(1)=0,
\end{equation}
where $R(\Tilde{\phi}):=\mu^{(n)}/\mu^{(s)}$ is the ratio of normal to suspension viscosity and $B_s= (\rho_p-\rho_\ell)\cot\alpha/\rho_\ell K_n\Lambda_2$ is a non-dimensional buoyancy parameter analogous to $B_d$ above. The solution of \eqref{eqn:SBM_eq} is unique if $\phi_{0}$ is given. Similar to the diffusive flux model, the first equation in \eqref{eqn:SBM_eq} is the leading order approximation of the Stokes equation, while the second equation approximates the particle transport equation in the thin film limit.

By comparing systems \eqref{eqn:dfm_ode} and \eqref{eqn:SBM_eq}, we observe two key points about these models in the thin-film limit. First, despite the suspension balance model using a stress tensor that more accurately describes the rheology of the particle-liquid mixture than the diffusive flux model, both models lead to the same reduced equation that relates shear stress and gravitational terms in the thin film limit. Second, the main difference between the two models is reflected in the second equations of \eqref{eqn:dfm_ode} and  \eqref{eqn:SBM_eq}. In the diffusive flux model, the reduced equation for particle flux depends only on the shear stress, whereas in the suspension balance model, it depends on both the shear stress and its derivative. This leads to different solution behaviors near the free surface.

Figure 5 in \cite{wong2019fast} shows the solution of \eqref{eqn:SBM_eq} for different values of $\phi_{0}$. When $\phi_0$ is large, $\phi = \phi_{m}$ for a region near the free surface, and this region expands as $\phi_0$ approaches the maximum packing fraction. This suggests a layer of particles at the maximum packing fraction suspended above the liquid and particle mixture, resulting in zero velocity in that region, which is non-physical. In contrast, the solution of \eqref{eqn:dfm_ode} avoids this issue as the solution only reaches the maximum packing fraction at a single point $s=1$. 
To address this problem, we follow the method proposed in \cite{wong2019fast} and include a regularization term $\dot{\gamma}_0$ in the shear rate, such that $\dot{\gamma} = \sqrt{|u_z|^2 + \dot{\gamma}_0^2}$. This term accounts for non-local effects relevant at a distance $\mathcal{O}(d)$ from the surface \cite{miller2006normal} and ensures the effective shear rate is not zero at the free surface. The corresponding modified equilibrium equations are then given by:
\begin{equation}\label{eqn:corrected_sbm_ode}
\partial_{s}(R(\tilde{\phi})\hat{\sigma}) + B_s\tilde{\phi}=0,\quad \hat{\sigma} = \sqrt{\tilde{\sigma}^2+\delta_0^2\mu^{(s)}(\tilde{\phi})^2},\quad \tilde\sigma(1)=0,
\end{equation}
where $\delta_0\ll1$ denotes a dimensionless $\dot{\gamma}_0$. With this modification, the equilibrium avoids reaching the maximum packing fraction within the domain. Fig.~\ref{fig:bifurcations}(b) shows an example for a fixed angle $\alpha = 50^\circ$.

Next, we want to find the film height as well as the locations of the particle and fluid fronts as the suspension moves down the incline. Substituting the approximation of the velocity field and particle volume fraction, and averaging in the $z$-direction, we obtain a conservation law system for $h (x,t)$ and depth-averaged particle concentration $n (x,t) =h (x,t) \smash{\int_{\vphantom{1} 0}^{1}}\tilde{\phi} (x,s,t) ds$,
\begin{equation} \label{eqn:conservation_law}
\partial_{t}h+\partial_{x}(h^3f(\phi_0))=0, \quad \partial_{t}n+\partial_{x}(h^3g(\phi_0))=0, 
\end{equation}
where the suspension and particle fluxes, $f$ and $g$, and the vertically averaged particle volume fraction $\phi_0$ are given by
\begin{align}\label{eqn:fluxes}
f(\phi_0)=\int_0^1\tilde{u}ds, \quad
g(\phi_0)=\int_0^1\tilde{\phi}\tilde{u}ds,\quad \phi_0=\int_0^1\tilde{\phi}ds=\frac{n(t,x)}{h(t,x)}.
\end{align}
The $2\times 2$ system of conservation laws \eqref{eqn:conservation_law} is valid for both models. Hence, the equilibrium and conservation law equations for the diffusive flux model are given by \eqref{eqn:dfm_ode}  and \eqref{eqn:conservation_law}, while those for the suspension balance model are given by  \eqref{eqn:corrected_sbm_ode} and \eqref{eqn:conservation_law}. In the next section, we compute numerical solutions to both models by solving the equilibrium equations \eqref{eqn:dfm_ode} and \eqref{eqn:corrected_sbm_ode}  and then using these solutions to compute the fluxes in the system \eqref{eqn:conservation_law}. As a reminder, the two models differ only in their formulation of the equilibrium model (\eqref{eqn:dfm_ode} and \eqref{eqn:corrected_sbm_ode}). 

\section{Numerical Simulations} \label{numeric sol}

Table \ref{tab:param_vals} records the numerical values of the characteristic scales and parameters we use in this section. The height scale is computed by $H=V_{i}/w_{i}L$, where $V_{i}$ denotes the initial volume of the suspension, which varies by experiment. 
\begin{table}[!ht]
  \centering
  \small 
\begin{tabular}{c|c|c|c|c|c|c|c|c} 
Parameter & $\rho_p$ & $\rho_\ell$ & $\mu_\ell$ & $g$ & $L$ & $\delta_0$ & $w_{i}$ & $d_{i}$ \\
\hline
Value & 2475 kg/m$^3$ & 971 kg/m$^3$ & 0.971 kg/(m$\cdot$s) & 9.8 m/s$^2$ & 1 m & $10^{-4}$ & 0.14 m & 0.10 m
\end{tabular}
  \caption{Parameters used for numerical computations in Section \ref{numeric sol}. $w_{i}$ and $d_{i}$ are the initial width and depth of the suspension, respectively. 
  }
  \label{tab:param_vals}
\end{table}
\subsection{Equilibrium Profile in $z$-direction }\label{equilibrium_solutions}

\begin{figure}[!ht]
    \centering
      \begin{subfigure}[h]{0.24\textwidth}
         \centering
         \includegraphics[width=\textwidth]{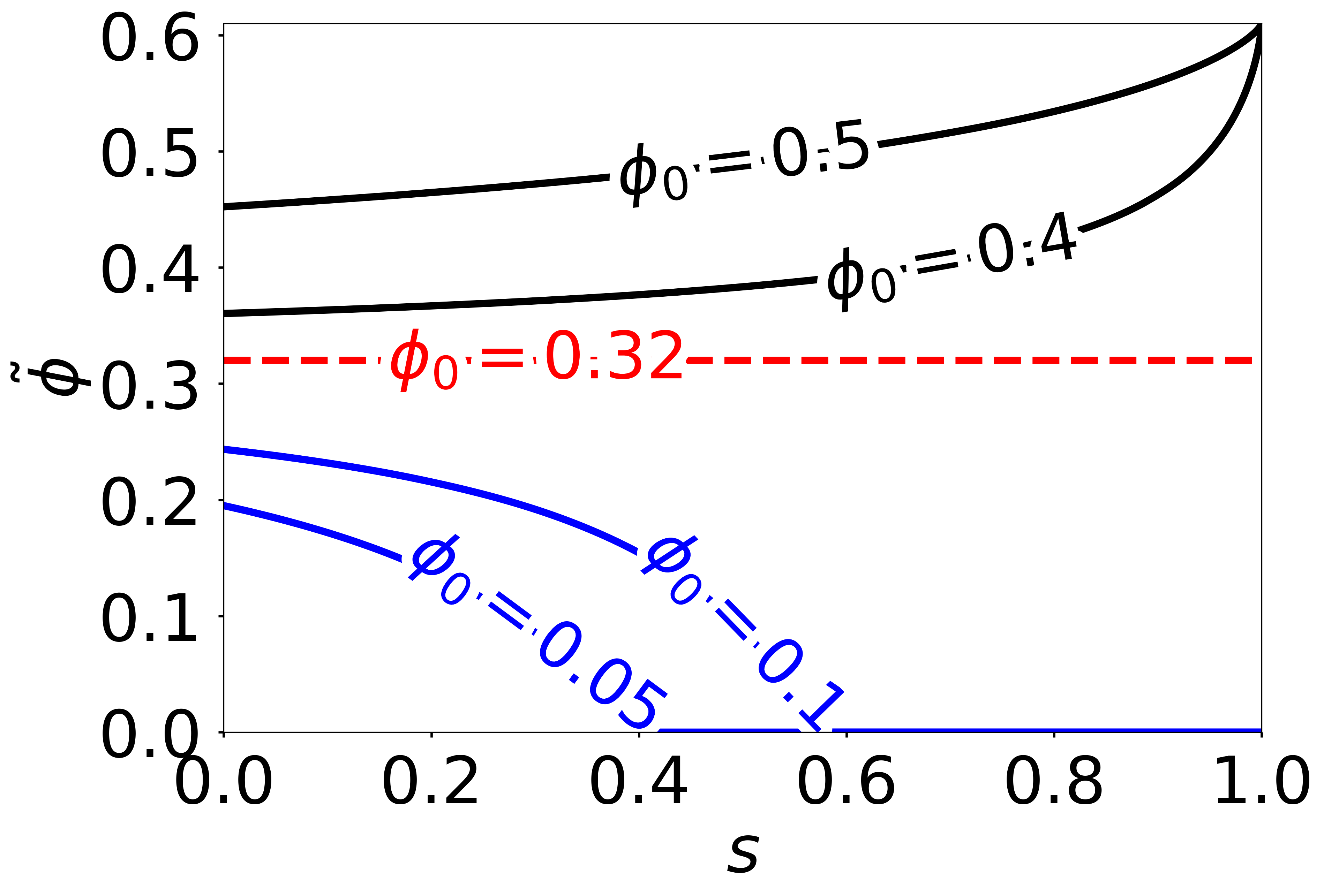}
         \caption{$\tilde{\phi}$, Diffusive flux model}
         \label{fig:phi_dfm_50}
     \end{subfigure}
     \begin{subfigure}[h]{0.24\textwidth}
         \centering
         \includegraphics[width=\textwidth]{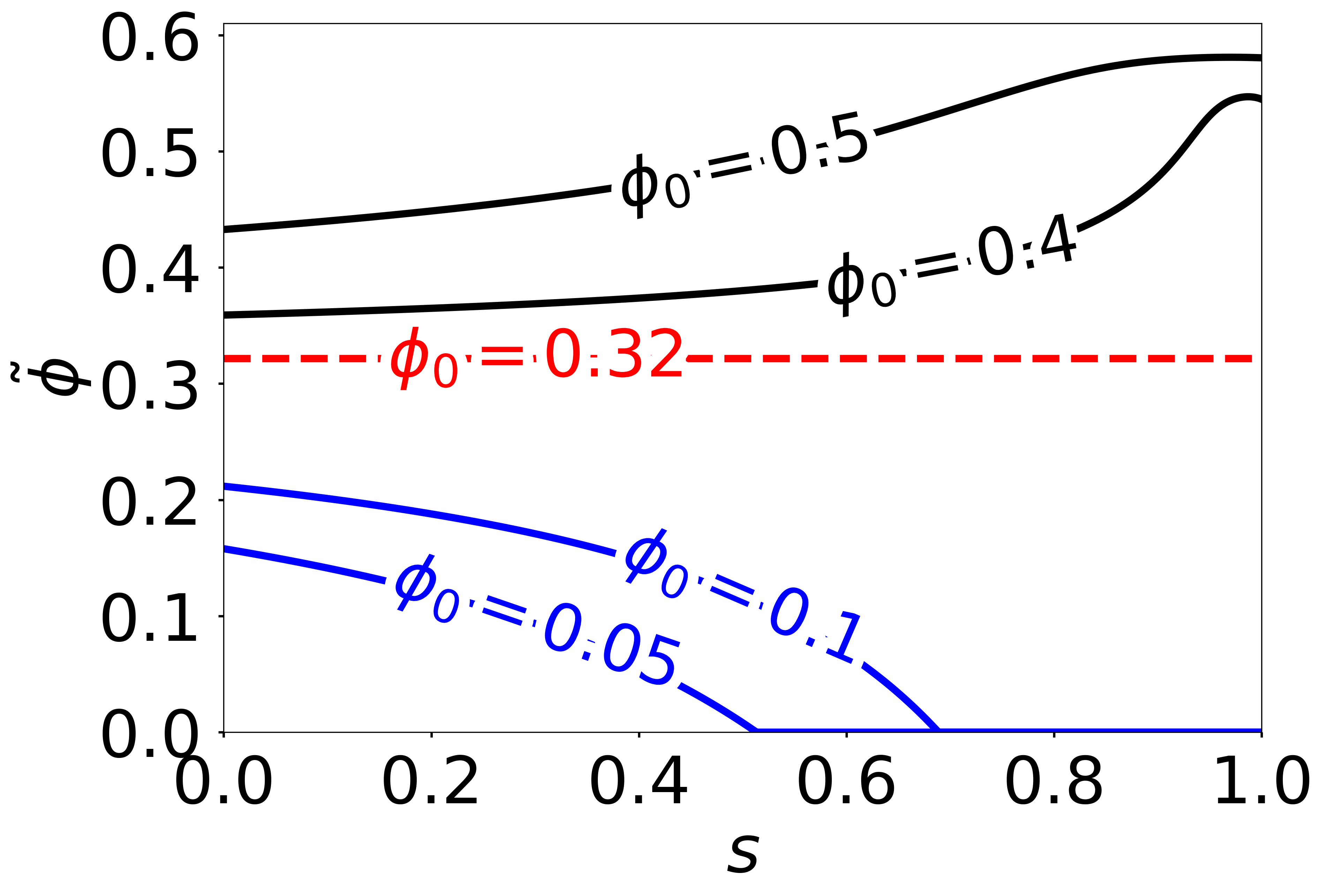}
         \caption{$\tilde{\phi}$, Suspension balance model}
         \label{fig:phi_sbm_50}
     \end{subfigure}
     \begin{subfigure}[h]{0.24\textwidth}
         \centering
         \includegraphics[width=\textwidth]{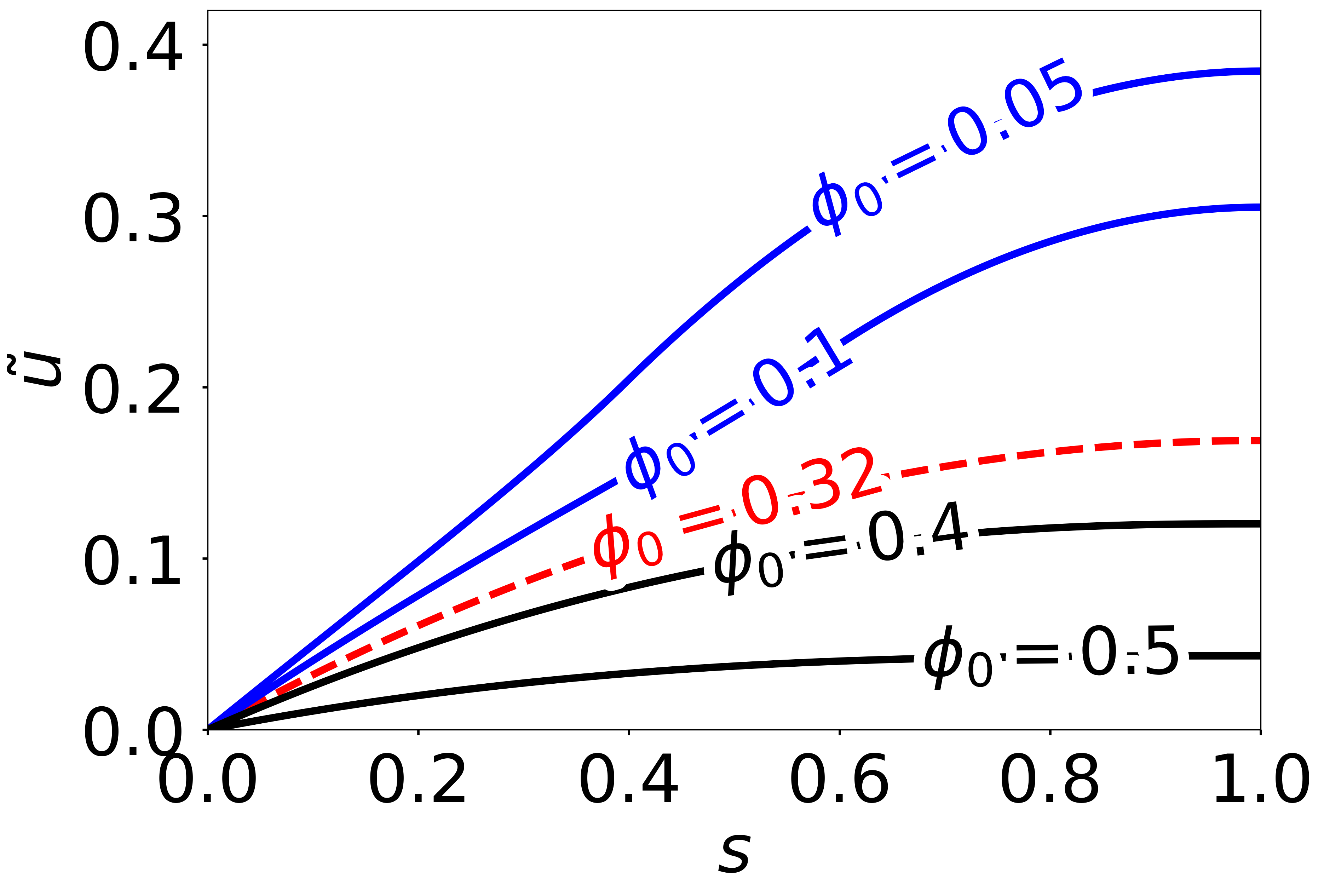}
         \caption{$\tilde{u}$, Diffusive flux model}
         \label{fig:u_dfm_50}
     \end{subfigure}
     \begin{subfigure}[h]{0.24\textwidth}
         \centering
         \includegraphics[width=\textwidth]{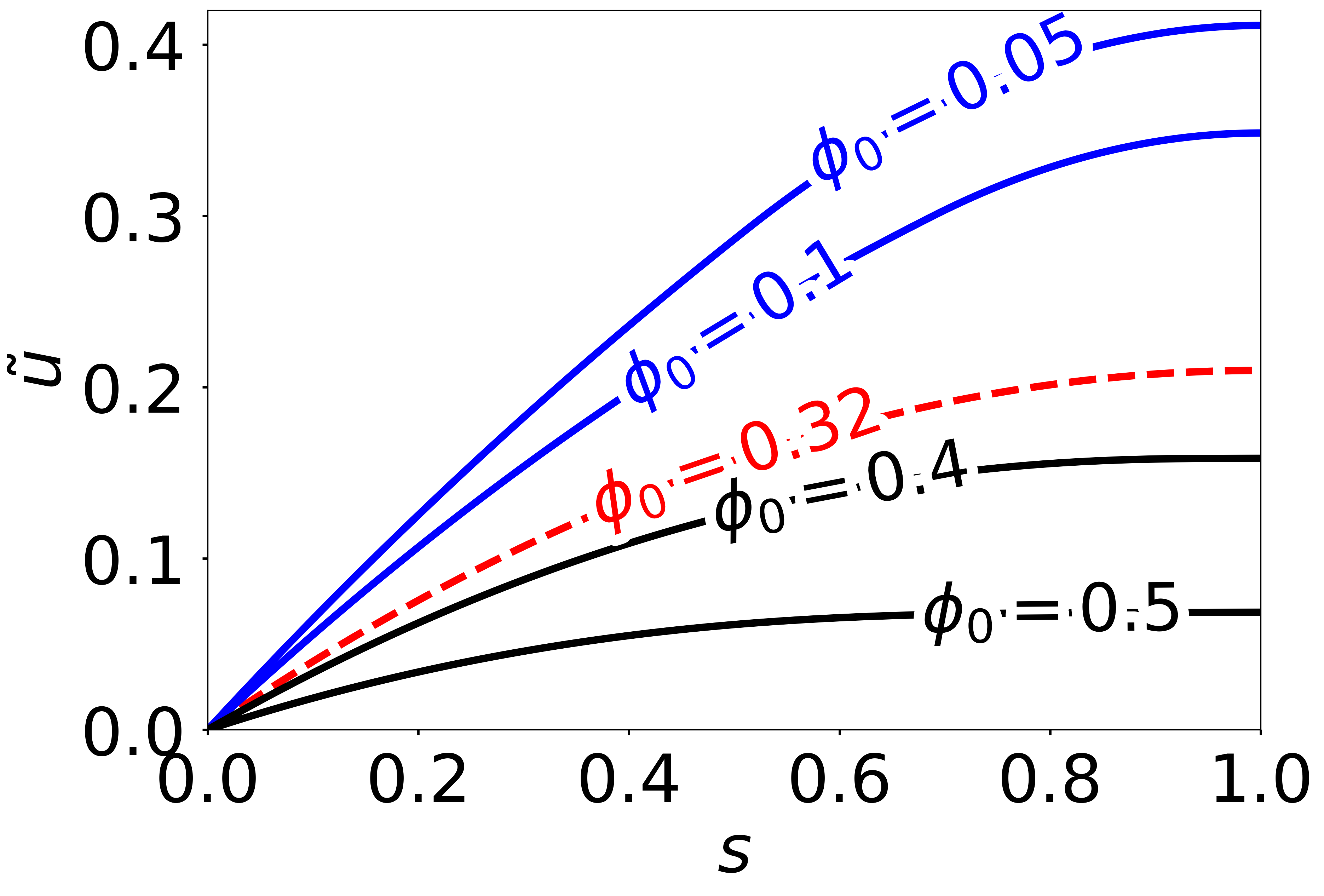}
         \caption{$\tilde{u}$, Suspension balance model}
         \label{fig:u_sbm_50}
     \end{subfigure}
        \caption{Particle volume fraction $\tilde{\phi}$ (\ref{fig:phi_dfm_50}, \ref{fig:phi_sbm_50}) and fluid velocity $\tilde{u}$ (\ref{fig:u_dfm_50}, \ref{fig:u_sbm_50}) profiles versus the non-dimensional height $s$ in the fluid layer for the diffusive flux and suspension balance models. The plots are generated at inclination angle $\alpha=50^\circ$ and maximum packing fraction $\phi_m=0.61$ to demonstrate different regimes. The families of solutions correspond to different $\phi_0$ values indicate by the labels on the plotted lines. The black lines correspond to the ridged regime, the red lines correspond to the well-mixed regime, and the blue lines correspond to the settled regime. Note that, for this particular value of \(\alpha\) and parameter values given in  Table \ref{tab:param_vals}, we find \(B_d=2.307\) and \(B_s=5.320\) for systems \eqref{eqn:dfm_ode} and \eqref{eqn:corrected_sbm_ode}.}
        \label{fig:bifurcations}
\end{figure}

Figure \ref{fig:bifurcations} shows a comparison of the diffusive flux and suspension balance models in terms of $\tilde{\phi}$ (\ref{fig:phi_dfm_50} and \ref{fig:phi_sbm_50}) and $\tilde{u}$ (\ref{fig:u_dfm_50} and \ref{fig:u_sbm_50}) generated from solutions to the equilibrium equations \eqref{eqn:dfm_ode} and \eqref{eqn:corrected_sbm_ode}. For all subfigures, we plot a family of solutions parametrized by $\phi_0$ at a fixed angle $\alpha=50^\circ$. For each solution, we plot $\tilde{\phi}$ as a function of $s$ and plot $\tilde{u}$ as a function of $s$. As reminder, $s = z/h$ is the non-dimensionalized fluid height, where $s=0$ is the bottom and $s=1$ is the fluid height.

For small $\phi_{0}$, $\tilde{\phi}$ has a larger value for small $s$ and is zero for large $s$, indicating that particles concentrate near the bottom, leaving clear fluid near the free surface. This is referred to as the settled regime can be seen in the blue curves of Fig.~\ref{fig:bifurcations}. As $\phi_{0}$ increases, the system transitions from the settled regime to the well-mixed regime (red curve), where the particles are uniformly distributed along the $z$-direction. We define such value of $\phi_{0}$ as the critical volume fraction, denoted as $\phi_{crit}$, since it is a bifurcation point of the system where the solutions with $\phi_{0}$ above and below $\phi_{crit}$  behave differently. With a further increase in \(\phi_{0}\), the system reaches the ridged regime (black curves), where particles are more concentrated near the free surface. Notice that, in the ridged regime, the two models have different behavior. As shown in Fig.~\ref{fig:phi_dfm_50}, the $\tilde{\phi}$ solutions of the diffusive flux model monotonically increase as $s$ increases. In contrast, on Fig.~\ref{fig:phi_sbm_50}, the $\tilde{\phi}$ profiles for the suspension balance model initially increase monotonically but then stabilize or even decrease slightly near $s=1$. This property results from the regularization term introduced in the modified shear rate. As discussed in Section \ref{theoretical_sol}, without the regularization term, $\tilde{\phi}$ would monotonically increase, reaching and remaining at $\phi_m$  for some point $s_{1}<1$ (see Figure 5 in \cite{wong2019fast}).

Figures \ref{fig:u_dfm_50} and \ref{fig:u_sbm_50} show velocity profiles for the same given $\phi_0$ values as the figures on the left. For both models, $\tilde{u}$ decreases as we increase $\phi_0$ and move from the settled to well-mixed and finally ridged regimes. This is expected, as higher particle concentrations increase the effective viscosity of the mixture, leading to reduced velocity. However, the suspension balance model on Fig.~\ref{fig:u_sbm_50} consistently predicts larger $\tilde{u}$ values than the diffusive flux model on Fig.~\ref{fig:u_dfm_50} at the same $\phi_0$.

\begin{figure}[!t]
     \centering
     \begin{subfigure}[h]{0.28\textwidth}
         \centering
         \includegraphics[width=\textwidth]{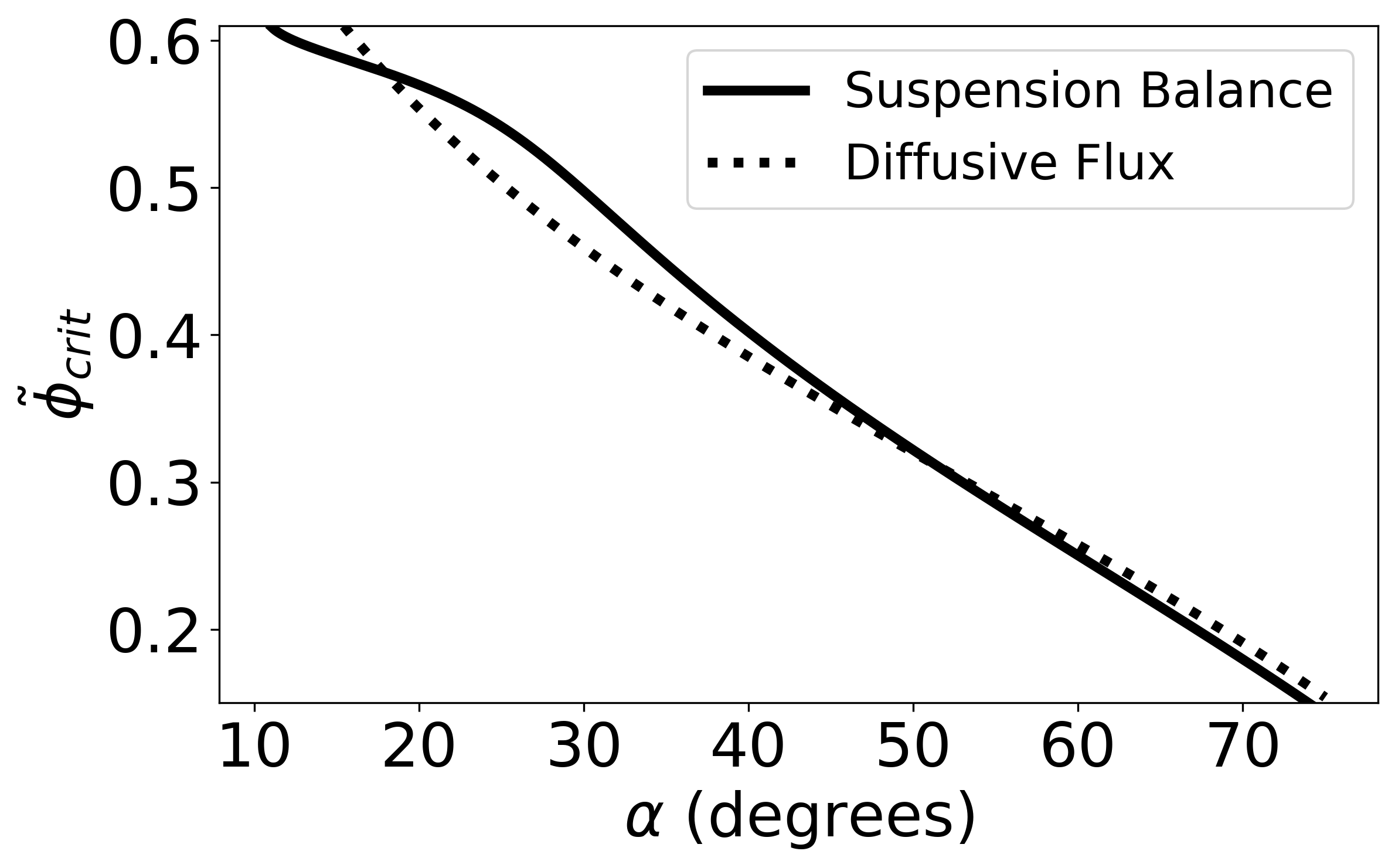}
         \caption{Critical volume fraction}
         \label{fig:phi_crit_61}
    \end{subfigure}
    \begin{subfigure}[h]{0.28\textwidth}
         \centering
         \includegraphics[width=\textwidth]{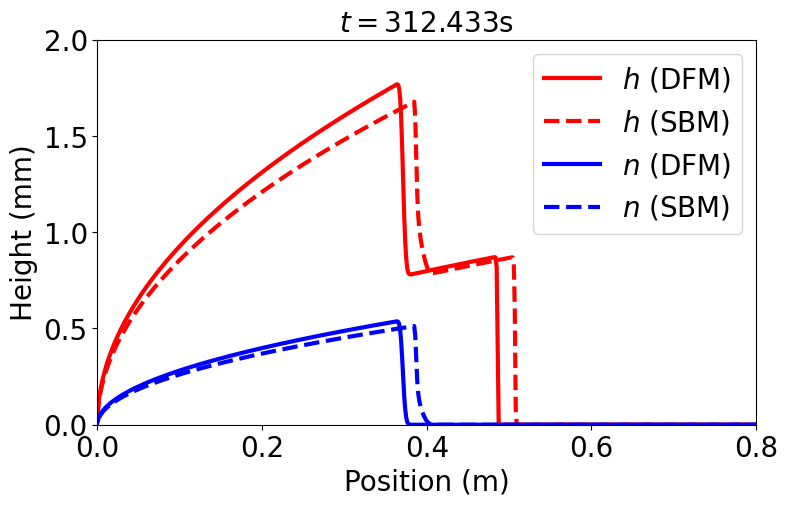}
         \caption{Free surface profile}
         \label{fig:height_1}
     \end{subfigure}
     \begin{subfigure}[h]{0.28\textwidth}
         \centering
         \includegraphics[width=\textwidth]{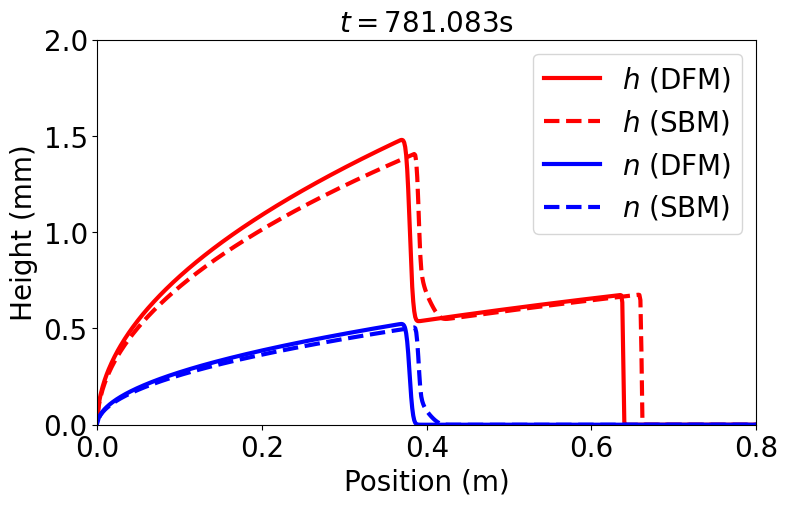}
         \caption{Free surface profile}
         \label{fig:height_2}
     \end{subfigure}
     \caption{(\ref{fig:phi_crit_61}) Critical volume fraction $\phi_{crit}$ for the two models when $\phi_m=0.61$. (\ref{fig:height_1}) \& (\ref{fig:height_2}) Height profiles $h$ (colored red) and depth-averaged particle concentration $n$ (colored blue) 
       for the two models at two different times. The parameters we use are $\phi_0=0.25$ and $\alpha=10^\circ$.}
    \label{fig:h_n_profiles}
\end{figure}

Since $\tilde{\phi}$ is a constant when $\phi_{0}=\phi_{crit}$, the critical volume fraction can be computed by setting $\partial \tilde{\phi}/\partial s=0$ and solving for $\tilde{\phi}$ separately using either \eqref{eqn:dfm_ode} and \eqref{eqn:corrected_sbm_ode}. We have
\begin{equation}\label{eqn:phi_crit}
    \tilde{\phi}_{crit,{DFM}}= \min\left\{\phi_m, \frac{-\rho_\ell(B_d+1)}{2(\rho_p-\rho_\ell)}+\sqrt{\left(\frac{\rho_\ell(B_d+1)}{2(\rho_p-\rho_\ell)}\right)^2+\frac{\rho_\ell B_d}{\rho_p-\rho_\ell}}\right\},\quad \tilde{\phi}_{crit,{SBM}}=\min\left\{\phi_m,\xi\right\},
\end{equation}
where $\xi$ is the unique solution to $-(1+(\rho_p-\rho_\ell)/\rho_\ell)R(\xi)+B_s=0$.
 Figure \ref{fig:phi_crit_61} illustrates the values of $\phi_{crit}$ as a function of inclination angle $\alpha$ for the DFM and SBM with $\phi_m=0.61$. This plot differs slightly from the comparison study in previous work (see Fig.~5(a) in \cite{wong2019fast}), where they used $\phi_m = 0.55$ and the regularization term from \eqref{eqn:corrected_sbm_ode} is not included. The $\phi_{crit}$ values for each model are similar for large angles. In fact, the maximum relative difference of $\phi_{crit}$ between the two models is only $6.715\%$. 
After obtaining the equilibrium solutions, we can compute the fluxes $f$ and $g$ in the conservation law \eqref{eqn:fluxes}. Then we can solve the conservation law to obtain the height profile and particle distribution at different time instances, which will be discussed in Section ~\ref{numerical_sol}.

\subsection{Evolution of Height Profiles, Fluid and Particle Fronts} \label{numerical_sol}
\begin{figure}[!t]
     \centering
     \begin{subfigure}[h]{0.23\textwidth}
         \centering
         \includegraphics[width=\textwidth]{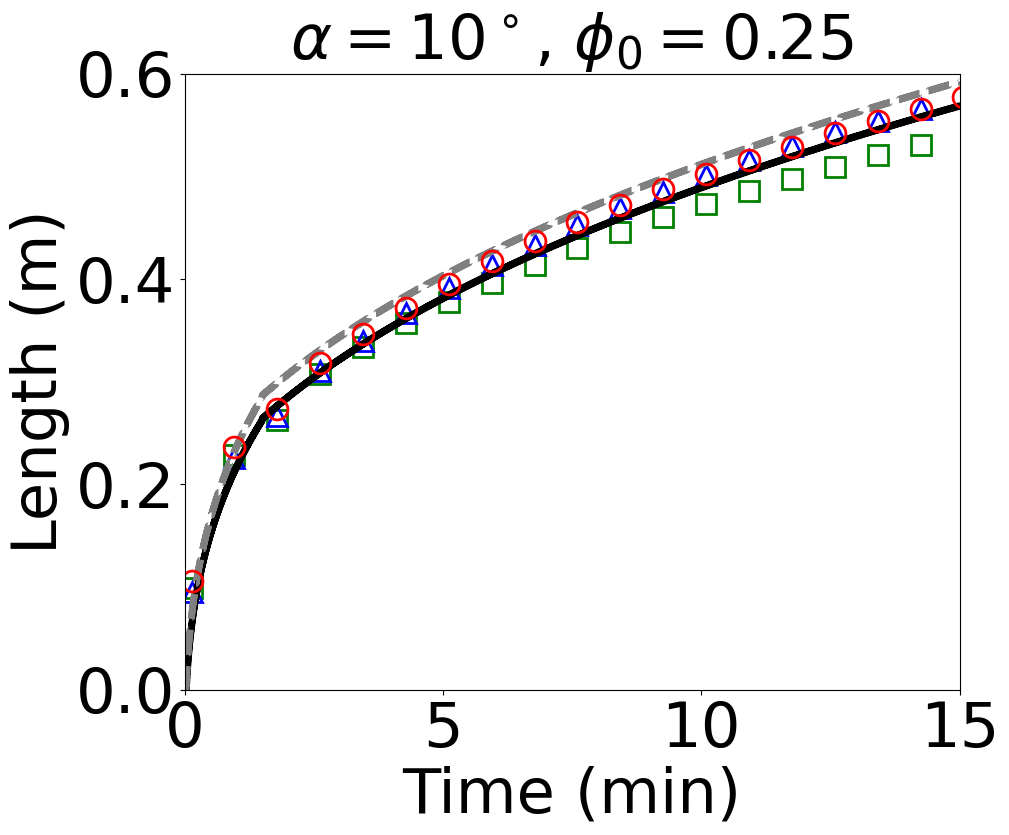}
         \caption{Liquid front}
         \label{fig:exp_cff}
     \end{subfigure}
     \begin{subfigure}[h]{0.23\textwidth}
         \centering
         \includegraphics[width=\textwidth]{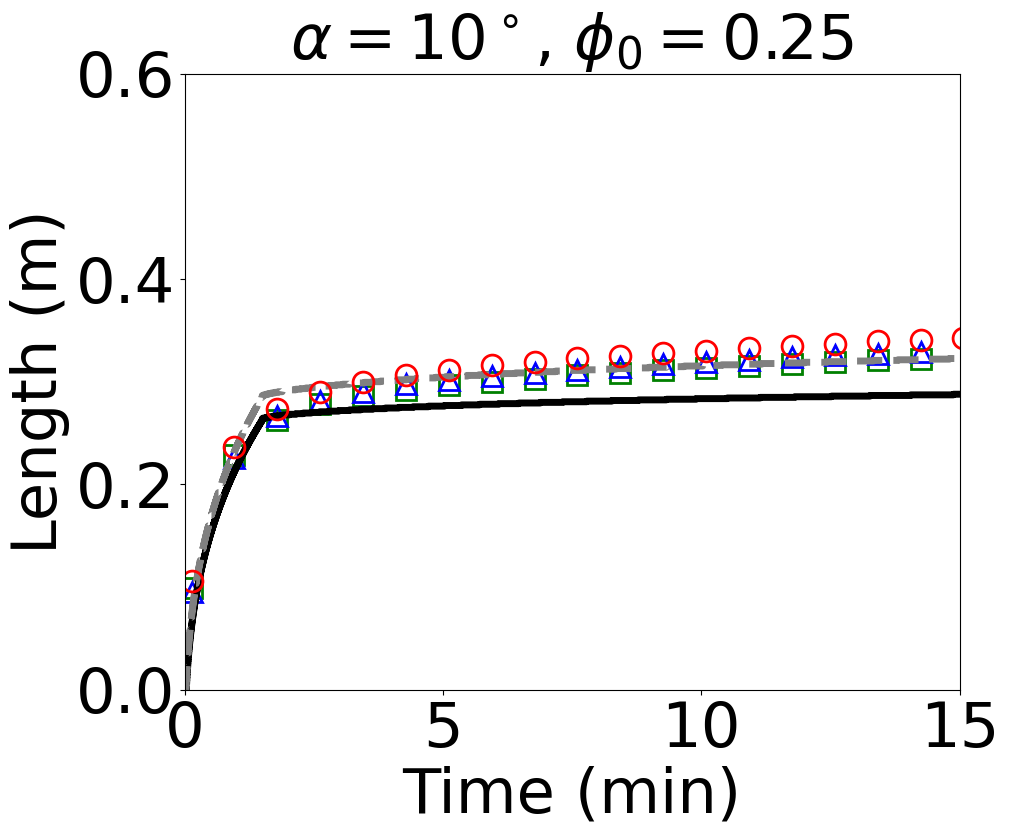}
         \caption{Particle front}
\label{fig:exp_pf}
     \end{subfigure}
        \label{fig:fronts}
     \begin{subfigure}[h]{0.23\textwidth}
         \centering
         \includegraphics[width=\textwidth]{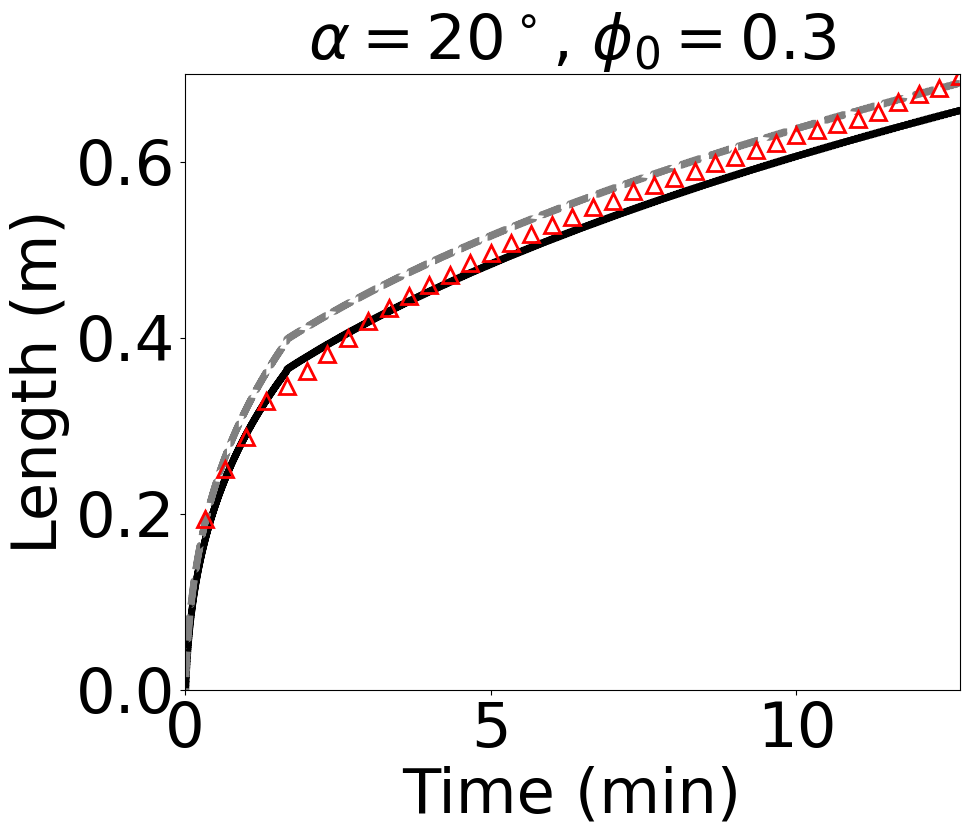}
         \caption{Liquid front}
         \label{fig:alpha=20_ff}
     \end{subfigure}
     \begin{subfigure}[h]{0.23\textwidth}
         \centering
         \includegraphics[width=\textwidth]{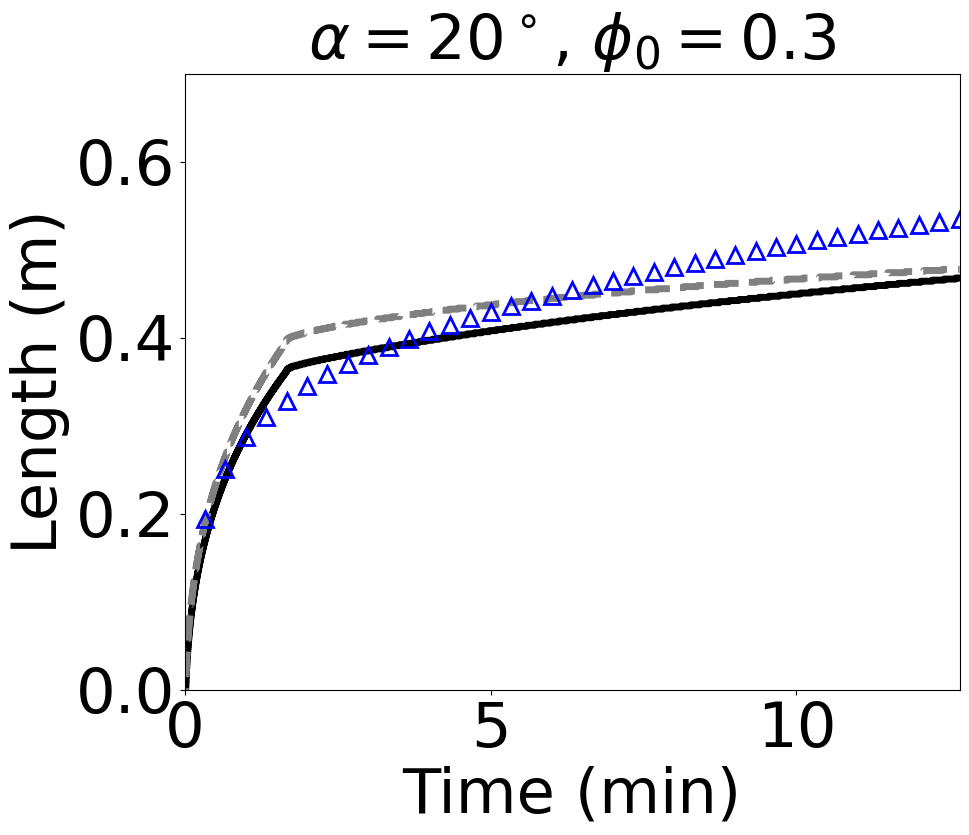}
         \caption{Particle front}
         \label{fig:alpha=20_pf}
     \end{subfigure}
     \caption{Liquid and particle front profiles with respect to time using parameters $\phi_0=0.25$ and $\alpha=10^\circ$ in \ref{fig:exp_cff} and \ref{fig:exp_pf}, and parameters $\phi_0=0.3$ and $\alpha=20^\circ$ in \ref{fig:alpha=20_ff} and \ref{fig:alpha=20_pf}. In each plot, numerical results are shown in black solid curves, which represent the DFM, and gray dashed curves for the SBM. Colored symbols indicate experimental results, with each color corresponding to a different experimental trial. Note that the numerical results include a transient regime as in \cite{murisic2013dynamics}. }
     \label{fig:liquid particle}
\end{figure}

Following \cite{murisic2013dynamics}, we verify that \eqref{eqn:conservation_law} is hyperbolic for the base parameters in Table \ref{tab:param_vals} and the angles $\alpha$ considered in Figures \ref{fig:height_1}, \ref{fig:height_2}, and \ref{fig:liquid particle}, and determine that the system is well-posed. We solve the system \eqref{eqn:conservation_law} using the first-order upwind scheme, which is simple and robust for solving hyperbolic systems. We obtain the initial conditions for \eqref{eqn:conservation_law} through the following process. As in \cite{murisic2013dynamics}, we first note that  \eqref{eqn:conservation_law} only holds after the end of a transient regime at $t_{trans}$, or once the particles have equilibrated in the $z$-direction. From actual experiments in the settled regime, we can estimate $t_{trans}$ by the time we see the particle and liquid fronts separate for the first time. Consequently, we focus on the settled regime in this section, as the other two regimes do not exhibit a separate liquid front, complicating experimental comparisons. 

Before $t_{trans}$, suspensions start well-mixed, the flow is unsteady, and the equilibrium assumption does not hold. We employ the following equation in Huppert \cite{huppert1982flow} to describe the dynamics of the transient regime:
\begin{align}\label{eqn:huppert}
h(t,x)=\left(\frac{\mu^{\ast}(\phi)}{(\rho_\ell+(\rho_p-\rho_\ell)\phi)g\sin\alpha}\right)^{1/2}\left(\frac{x}{t}\right)^{1/2},\quad 0\leq x\leq\left(\frac{9A^2g\sin\alpha(\rho_\ell+(\rho_p-\rho_\ell)\phi)}{4\mu^{\ast}(\phi)}\right)^{1/3}t^{1/3}.
\end{align}
 $\mu^{\ast}(\phi)$ is the viscosity of the suspension, which is the Krieger-Dougherty relation for the DFM and \eqref{eqn:sbm_visc} for the SBM. Moreover, $A=V_{i}/w_{i}$ is the initial cross-sectional area of the suspension. Using the formula for $h$ given in \eqref{eqn:huppert}, we then have $n=h\phi_0$ as we assume that the suspension is well-mixed during this transient regime. With analytical expressions for $h$ and $n$ in the transient regime, we can solve for the profiles at $t_{trans}$, and use the results at that time as the initial conditions for \eqref{eqn:conservation_law}. This is the procedure described in \cite{murisic2013dynamics}. 

We compute particle and liquid front positions as a function of time, respectively, by determining where the solution profiles fall below a certain threshold (e.g. Figure~\ref{fig:liquid particle}). 
Figures \ref{fig:height_1} and \ref{fig:height_2} compare the height profiles $h$ and depth-averaged particle concentration $n$ of the solutions to the two models. The DFM yields similar solutions when compared with the SBM.
We observe that as both $h$ and $n$ evolve from Figure \ref{fig:height_1} to Figure \ref{fig:height_2}, there is a significant forward movement of the clear fluid front (red curves), compared to the particle front (blue curves), which qualitatively describes the experimental observation.

Next, we compare the positions of the liquid and particle fronts. In the settled regime, the distances of the fluid and particle fronts in the $x$-direction (averaged in the $y$-direction) are easily measurable, making them ideal for temporal comparison. Figs. \ref{fig:exp_cff} and \ref{fig:exp_pf} compare liquid and particle fronts from numerical simulations with physical experiments, which systematically follow the procedures described in \cite{murisic2013dynamics}. In particular, we experimentally measure that the percent loss of the suspension on the reservoir gate and in the cup is around $25\%$ on average, resulting in $V_{i}=75\text{mL}$ and assume that the suspension is initially well-mixed. Our experimental estimate for $t_{trans}$ is $90\text{s}$. Figs.~\ref{fig:alpha=20_ff} and \ref{fig:alpha=20_pf}
are comparisons of numerical liquid and particle fronts against another experiment with a steeper angle of inclination and higher initial concentration, where we have $V_{i}=82.5\text{mL}$ and $t_{trans}=100$s.

We compute the average maximum relative differences between the numerical and experimental profiles for both fronts, both models, and both sets of parameters. From Fig.~\ref{fig:liquid particle}, the SBM  fronts move faster than the DFM fronts. The DFM fits the experimental liquid fronts better for both sets of parameters, with an average maximum relative difference of 4\% to 6\%. The SBM fits the experimental particle fronts better for \(\alpha=10^\circ\) and \(\phi_0=0.25\), achieving a difference of 9\%. Overall, though, the two models are quite similar, with differences generally less than or equal to 15\%. Only Fig. \ref{fig:alpha=20_pf} is related to larger differences, up to about 17\% for the suspension balance model, as the numerical particle fronts first overshoot and then undershoot. However, this result is still consistent with those of \cite{murisic2013dynamics}, particularly Fig. 12c in their paper.


In addition, we compute the maximum relative difference between the front positions of the two models over the whole simulation with the parameter $V_{i}=75\text{mL}$, $t_{trans}=90$s, and the nondimensional timestep size to be $1/2^{16}\approx 1.526\times 10^{-5}$, and the evolution time to be $900$s. Overall, we observe that all relative differences between  are sufficiently small, ranging from around $6\%$ to $9\%$. Given noises across experiments and nonuniform fingers typically present for the liquid fronts, we see that the two models agree quantitatively to a reasonable extent, at least in the settled regime.

\section{Conclusion}\label{conclusion}

The numerical solutions to both the DFM and the SBM
have similar liquid/particle front positions and height profiles, showing a good agreement with experimental data. Also, neither model is computationally expensive: on a desktop computer, sufficient accuracy (as demonstrated in Figs.~\ref{fig:h_n_profiles} and \ref{fig:liquid particle}) can be achieved with a first-order scheme for both methods with a runtime of approximately 20-30 minutes. We conclude that both models are similarly effective in the settled regime, and the differences of a single normal stress term and a particle flux expression in equations \eqref{eqn:shear_rate} and \eqref{eqn:particle_flux_sbm} are insignificant. 

There are several interesting avenues of investigation based on our work. One question is whether the two models agree for non-flat substrates. Moreover, the study of the effectiveness of the two models for higher volume fraction (in particular, in the ridged regime) is worth further investigation. 

\section*{Data Availability}

A Github repository will be made available once the paper is accepted for publication. 

\section*{Acknowledgements}

This material is based upon work supported by the U.S. National Science Foundation under
award No. DMS-2407006. 
This work is also supported by Simons Math + X Investigator Award number 510776.
Sarah C. Burnett was supported by the 2022 L'Or\'eal USA for Women in Science Postdoctoral Fellowship. We would also like to thank graduate students Hong Kiat Tan, Evan Davis, and Jack Luong for their helpful comments on the manuscript.

\bibliographystyle{elsarticle-harv}

\end{document}